\documentclass[preprint,12pt]{elsarticle}

\usepackage{graphicx,xcolor}
\usepackage{amssymb}

\usepackage{lineno}

\journal{Sensors and Actuators A: Physical}

\begin{document}

\begin{frontmatter}


\title{Large $d_{33}$ Piezoelectric-Polymer Composites For RF Acoustic Resonators}



\author{Pallabi~Das}
\author{Siddharth~Tallur}

\address{Department of Electrical Engineering, IIT Bombay,
              Mumbai 400076, Maharashtra, India}

\begin{abstract}
While piezoelectric transduction enables designing acoustic resonators operating at multi-GHz frequencies, the deposition of piezoelectric materials typically requires high temperature processes and specific crystallographic orientation of substrates, thus imposing a limitation on materials that could be used. In this paper we present a piezoelectrically transduced thickness mode acoustic resonator that employs piezoelectric (PMNPT) nanoparticles embedded in a polymer (SU8) matrix. This composite material is deposited using standard resist-spin coaters and is thus compatible with a variety of substrates. The device presented here uses a double side polished single crystal silicon wafer as the low loss acoustic substrate for the resonator and $1.7\mu m$ thick SU8-PMNPT composite film as the actuator, and exhibits large effective piezoelectric coefficient $(d_{33})$ of $216pm/V$, and we experimentally demonstrate efficient transduction of acoustic resonances at frequencies up to $1.5GHz$.
\end{abstract}


\end{frontmatter}


\section{Introduction}
\label{intro}
\begin{figure}[htbp]
		\centering
		\includegraphics[width= 9cm]{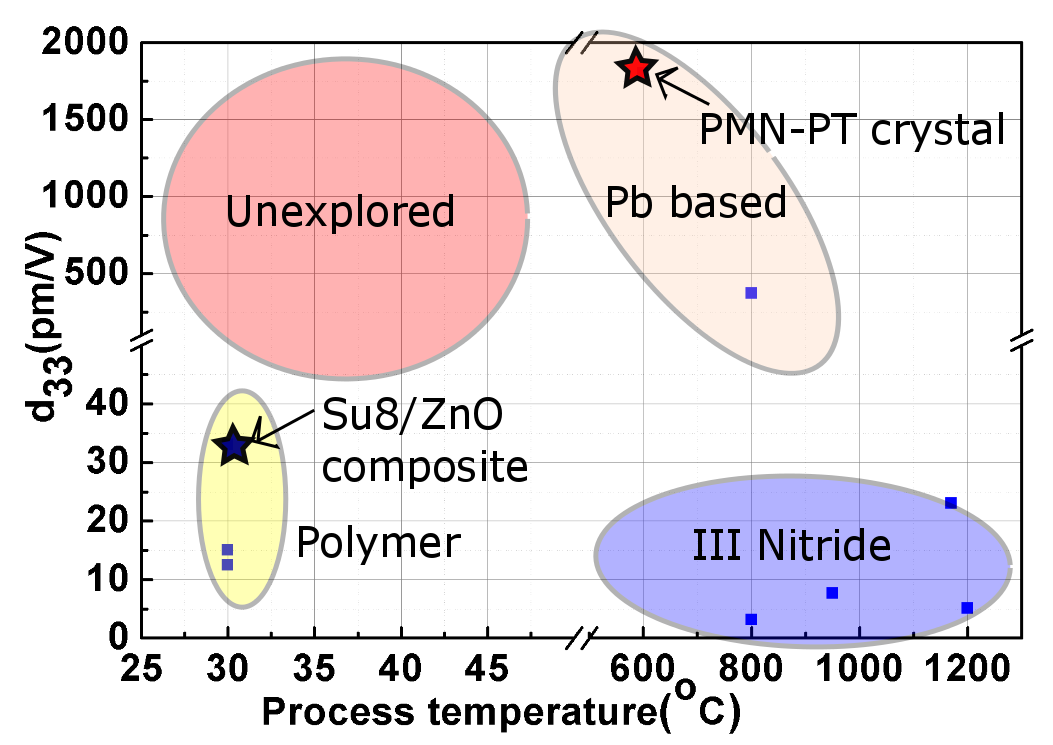}
		\caption{Piezoelectric coefficient and processing temperatures of conventional piezoelectric materials. This study provides a solution to the hitherto unexplored quadrant of a material that can be processed at ambient temperatures, with large piezoelectric coefficient.}
		\label{process_temp}
	\end{figure}

With the advent of the internet of things, MEMS technology and sensor miniaturization have seen unprecedented surge in interest. Piezoelectric transduction is widely employed in MEMS devices largely due to high transduction efficiency compared to electrostatic capacitive transduction. The resonators chosen for the purpose of this study are thickness mode resonators, that find applications in several areas such as low loss RF filters \cite{filter}, acoustic stress generators \cite{tanay}, frequency stabilization \cite{oscillators}, bio-sensing \cite{biosensing} and several other sensor applications \cite{appl}. High Overtone Bulk Acoustic Resonators (HBARs) are commmonly used thickness mode resonators, and offer the ability to have discrete tuning capability thanks to multiple acoustic resonances, with a fixed frequency spacing (defined by piezoelectric layer thickness) spread over a wide frequency range. The HBAR consists of a stack defined by a piezoelectric material sandwiched between thin metal electrodes. This stack is mechanically coupled to a low loss acoustic substrate that supports higher order harmonics of acoustic standing waves that form the resonances. The piezoelectric film thickness equals one-half acoustic wavelength in such a configuration. The transduction efficiency of the acoustic waves depends greatly on the electro-mechanical properties of the piezoelectric film.
The transduction efficiency of piezoelectric materials is characterized by the piezoelectric coefficient $(d_{33})$, which is a key parameter that directly impacts the electromechanical coupling efficiency, thus determining sensor performance. Most piezoelectric materials with large $d_{33}$ are lead based crystals such as lead zirconate titanate (PZT) or lead magnesium niobate-lead titanate (PMNPT), that typically require very high temperature $(>600^oC)$ processing for deposition on to substrates \cite{pzt}. These materials also require high temperature sol-gel deposition techniques or other specialized equipment that makes them incompatible with CMOS technology. On the other hand, popular materials such as aluminum nitride (AlN) and gallium nitride (GaN) that are easier to handle from a surface micro-machining perspective also typically require high processing temperatures but have lower piezoelectric coefficients \cite{nitride1},\cite{nitride2}. Ambient temperature RF magnetron sputtering can be used to deposit zinc oxide (ZnO) for piezoelectric transducers \cite{tanay}, but like the III-nitrides, the quality of the film growth on silicon or sapphire wafers also requires specific seed layers and crystallographic orientation of the substrate. Some groups have tried to address this challenge by developing acoustic resonators with room temperature processing using polymer piezoelectric materials such as PVDF \cite{pvdf} or composites \cite{su8},\cite{d33_cal} where the film quality is substrate-independent, however the measured $d_{33}$ has been reported to be very low. Figure~\ref{process_temp} pictorially shows the technology gap in availability of large $d_{33}$ piezoelectric material that requires low temperature processing.

In this article we present detailed methodology of fabricating a piezoelectric (PMNPT)-polymer (SU8) composite that can be deposited on any substrate via spin coating. This CMOS compatible process makes it possible to integrate the piezoelectric composite material with any substrate at room temperature. The following section provides details on preparation of the composite and resonator fabrication process flow. Subsequent sections highlight the material characterization and experimental characterization results of the resonators.
We observe acoustic resonances at frequencies up to $1.5GHz$, and measure the highest effective $d_{33}$ on the sample to be $216pm/V$.
While our work uses SU8 as the polymer due to its ease of handling and processing with photolithography, any polymer that can be deposited via spin-coating could be used for realizing devices such as the ones presented in this work.

 

\section{Device fabrication}

\subsection{Preparation of SU8/PMNPT composite}
\begin{figure}[htbp]
		\centering
		\includegraphics[width= 9cm]{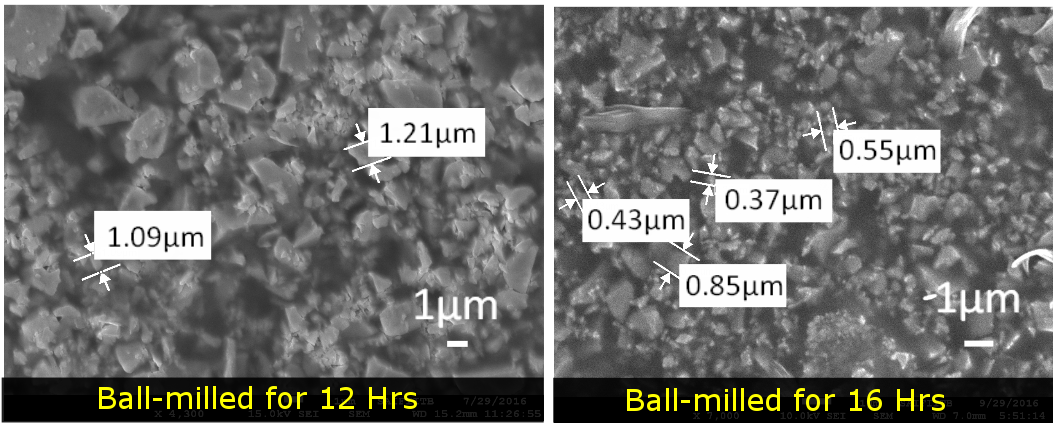}
		\caption{Impact of ball milling duration on the size of PMN-PT nanoparticles. 16 hours of ball milling results in nano-sized particles that can be uniformly dispersed in the polymer matrix.}
		\label{ballmill_sem}
	\end{figure}

Commercially available bulk crystals of PMNPT are mixed with cyclopentanone (SU8 thinner) and ground in a ball-miller for 16 hours to facilitate uniform dispersion of the resultant PMNPT nanoparticles in the solvent. The size of the resultant nanoparticles is predominantly determined by the duration of ball-milling. The SEM images shown in Figure \ref{ballmill_sem} show the nanoparticle size following 12 hours and 16 hours of ball-milling. The dispersed solution is then probe sonicated with SU8 2002 to prepare the PMNPT/SU8 composite. Instead of using commercial PMNPT crystals, one may also use PMNPT powder for the preparation of SU8/PMNPT composite. Synthesis of relaxor $0.68$PMN-$0.32$PT nano-powders in-house at low temperature is challenging, and therefore we have used commercially available crystals with high piezoelectric coefficient for our experiment. While large PMNPT coverage on the wafer is desirable, a high molar concentration of PMN-PT makes it difficult to pattern the composite film using photolithography. Through controlled characterization of the effectiveness of lithography, an optimum value of $7\%$ by volume of PMNPT in SU8 was obtained.

\subsection{Fabrication process flow}
\begin{figure}[htbp]
	\centering
	\includegraphics[width= 3.4in]{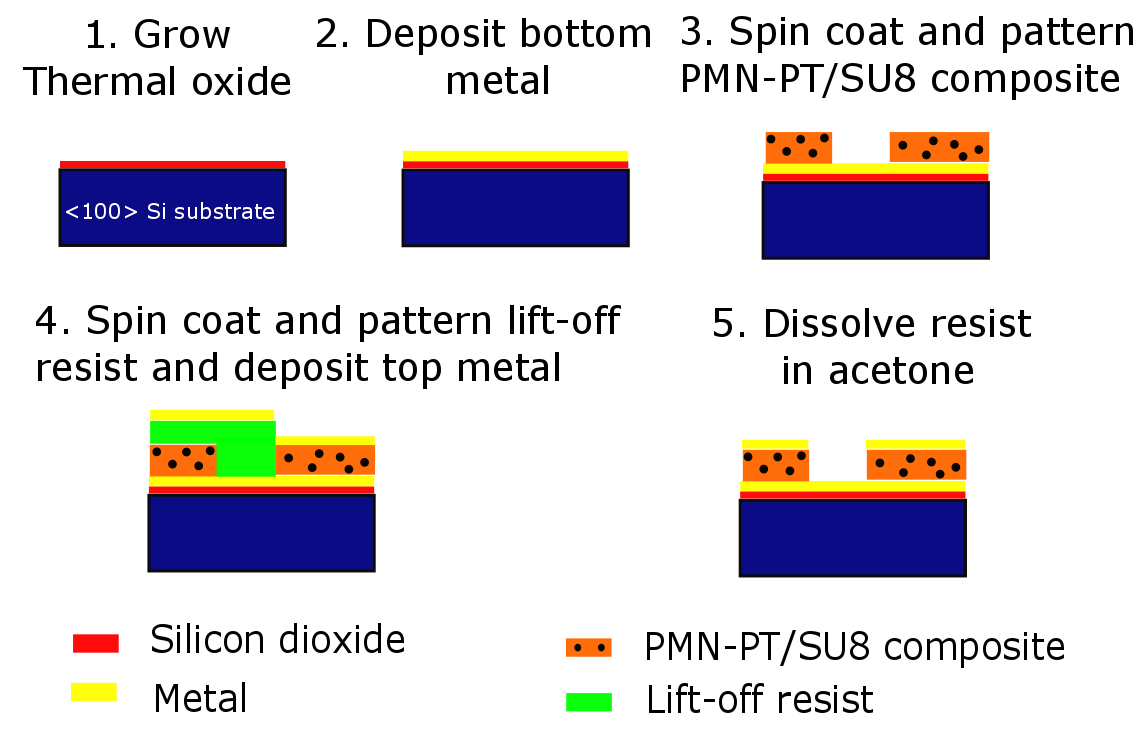}
	\caption{Illustration of resonator fabrication process flow using SU8/PMN-PT composite film as the piezoelectric layer.}
	\label{cross_sec}
\end{figure}

Figure~\ref{cross_sec} shows an illustration of the fabrication process flow. The first step involves thermal oxidation of $<$100$>$ oriented double-side polished, high resistivity silicon wafer (2 inch diameter, 275$\mu m$ thick) to obtain $200nm$ silicon dioxide to minimize substrate leakage. The bottom metal stack of chromium $(20nm)$/gold $(200nm)$ is then deposited via thermal sputtering. The roughness of the metal surface due to sputtering aids the adhesion of the SU8 polymer with the metal. In addition to this, we spin coat the bottom metal surface with Omnicoat to improve adhesion of the polymer composite with gold. Next the PMNPT/SU8 composite is spin coated and patterned using photolithography. Spin parameters corresponding to a $1.7\mu m$ thick composite layer are used to deposit the film. A thinner layer is required for higher frequency devices, but the film thickness was eventually pragmatically determined based on the size of the nanoparticles. The top metal comprising of a stack of titanium $(15nm)$/platinum $(150nm)$ is sputtered and patterned with a second photolithography mask via lift-off.

\begin{figure}[htbp!]
 	\centering
 	\includegraphics[width= 3.4in]{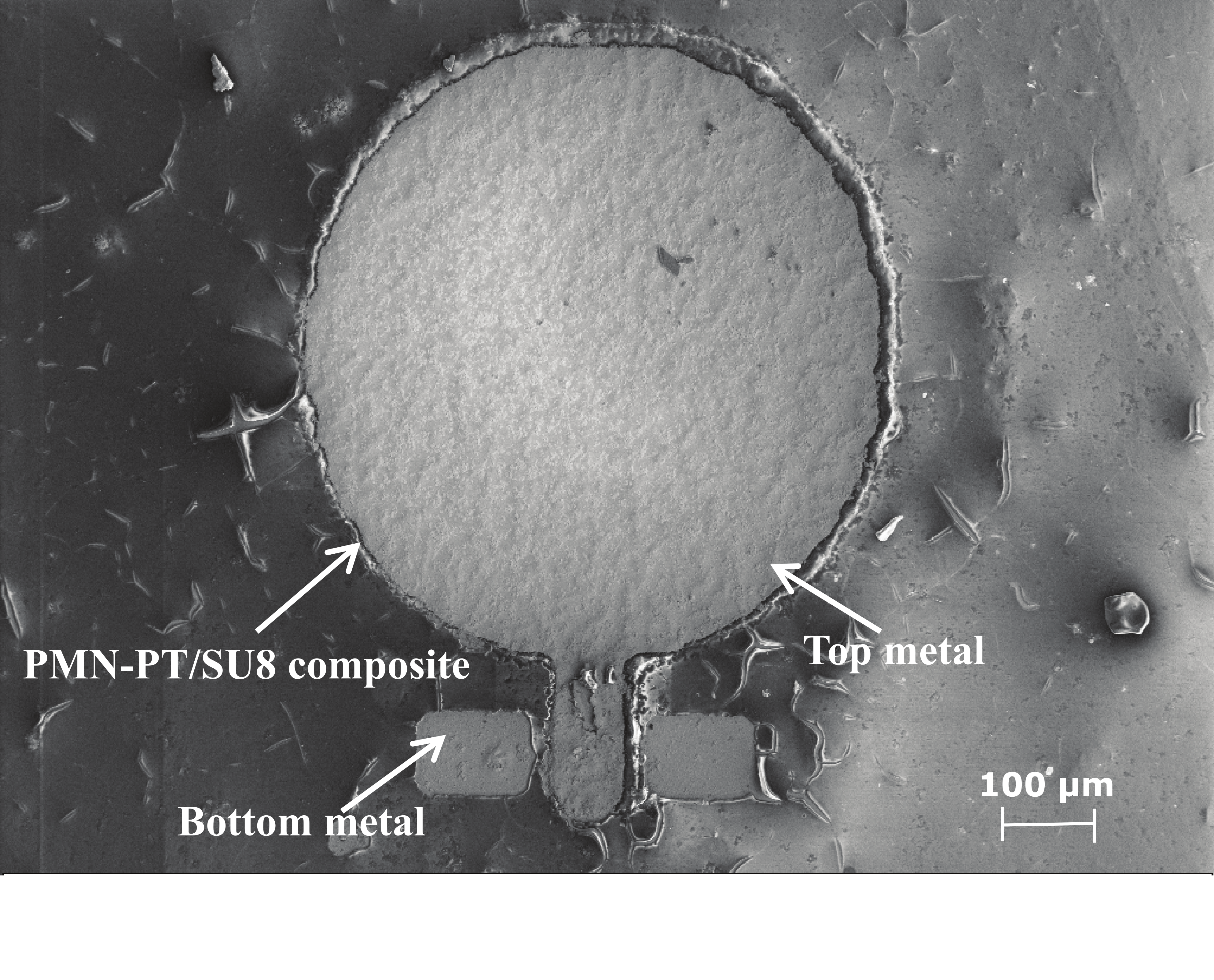}
 	\caption{Scanning electron micrograph (SEM) of the fabricated HBAR device with circular geometry. The roughness of the bottom metal is evident in the SEM.}
 	\label{Hbar_sem}
 \end{figure}

The resonator geometry comprises of circular shape patterns of the piezoelectric film of three different radii $(250\mu m,300 \mu m,350 \mu m)$, fabricated on a low loss $<$$100$$>$ silicon wafer. The bottom electrode (Cr/Au) is not patterned. The piezoelectric film and the top electrode (Ti/Pt) are patterned in circular shapes, with the top electrode diameter recessed by $10\mu m$ to avoid electrical shorting between top and bottom metals due to mask misalignment. Figure \ref{Hbar_sem} shows the SEM image of a fabricated device with radius $250 \mu m$.

\section{Experimental results}
\subsection{Film characterization}

\begin{figure}[htbp!]
  	\centering
  	\includegraphics[width= 9cm]{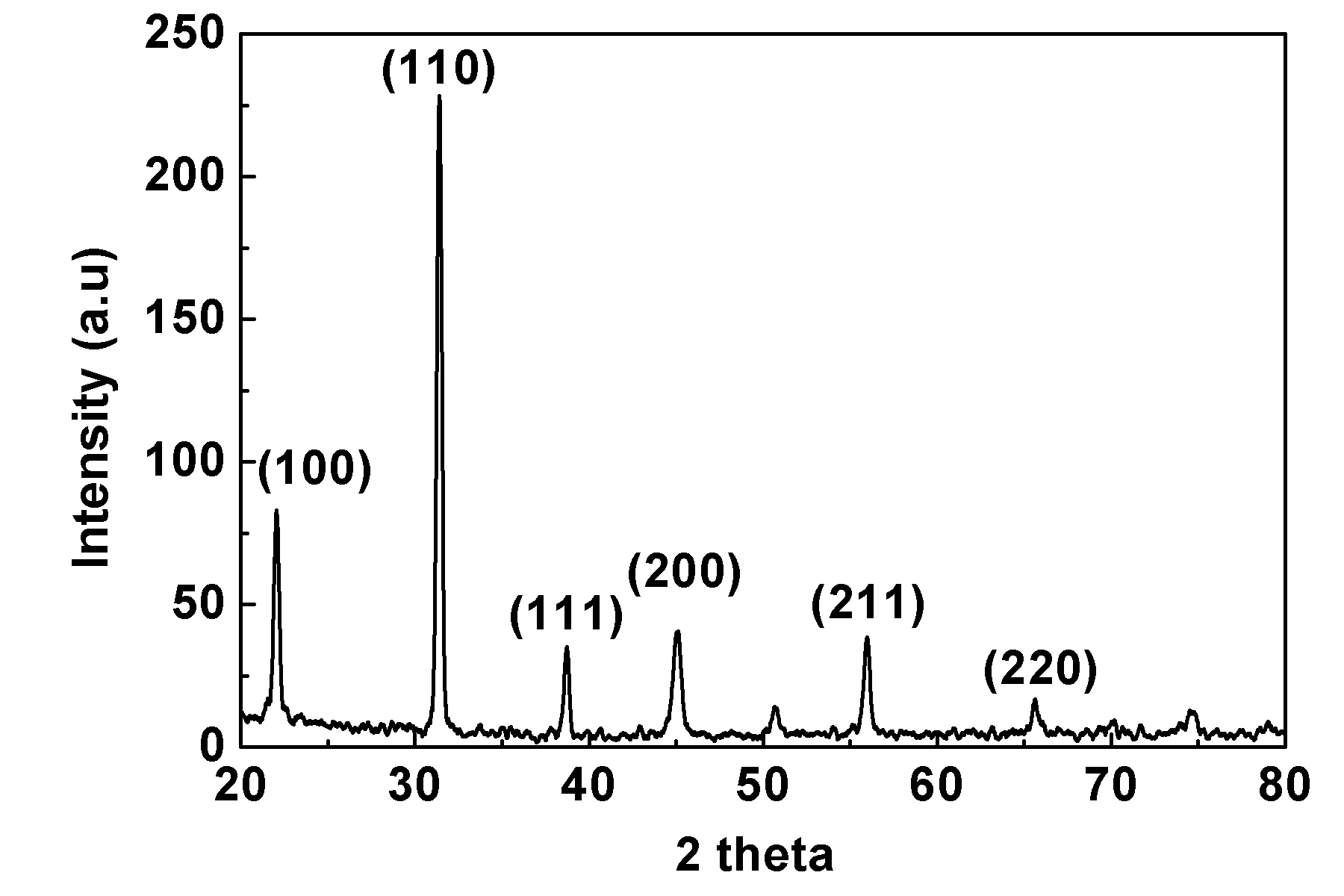}
  	\caption{X-Ray Diffraction (XRD) spectrum of the composite, showing that the polycrystalline nature of the nanoparticles is preserved after ball milling.}
  	\label{xrd_comp}
  \end{figure}

We obtain X-Ray Diffraction (XRD) spectrum of the composite, shown in Figure \ref{xrd_comp}. From this figure, it is evident that the polycrystalline property of the piezoelectric material is retained after ball milling. Since PMNPT is a ferroelectric material, it is possible to increase the $d_{33}$ by domain engineering through poling in a constant electric field. Due to the brittle nature of the polymer film, we cannot use conventional ferroelectric hysteresis and $d_{33}$ measurement setups for our device. Poling of the sample as well as measurement of effective $d_{33}$ is hence performed via Piezoresponse Force Microscopy (PFM) using Asylum MFP3D Origin conductive Atomic Force Microscope (c-AFM), as presented in other work in literature on lithium niobate resonators \cite{ryan}. Kvasov et al. \cite{ncomm} have reported such a technique for measurement of effective $d_{33}$ for PZT nanowires. Readers interested in further understanding the accuracy of PFM measurements for effective $d_{33}$ are encouraged to read an analysis of this technique published by Kvasov et al. \cite{ncomm}.

\begin{figure}[htbp!]
   	\centering
   	\includegraphics[width=7cm]{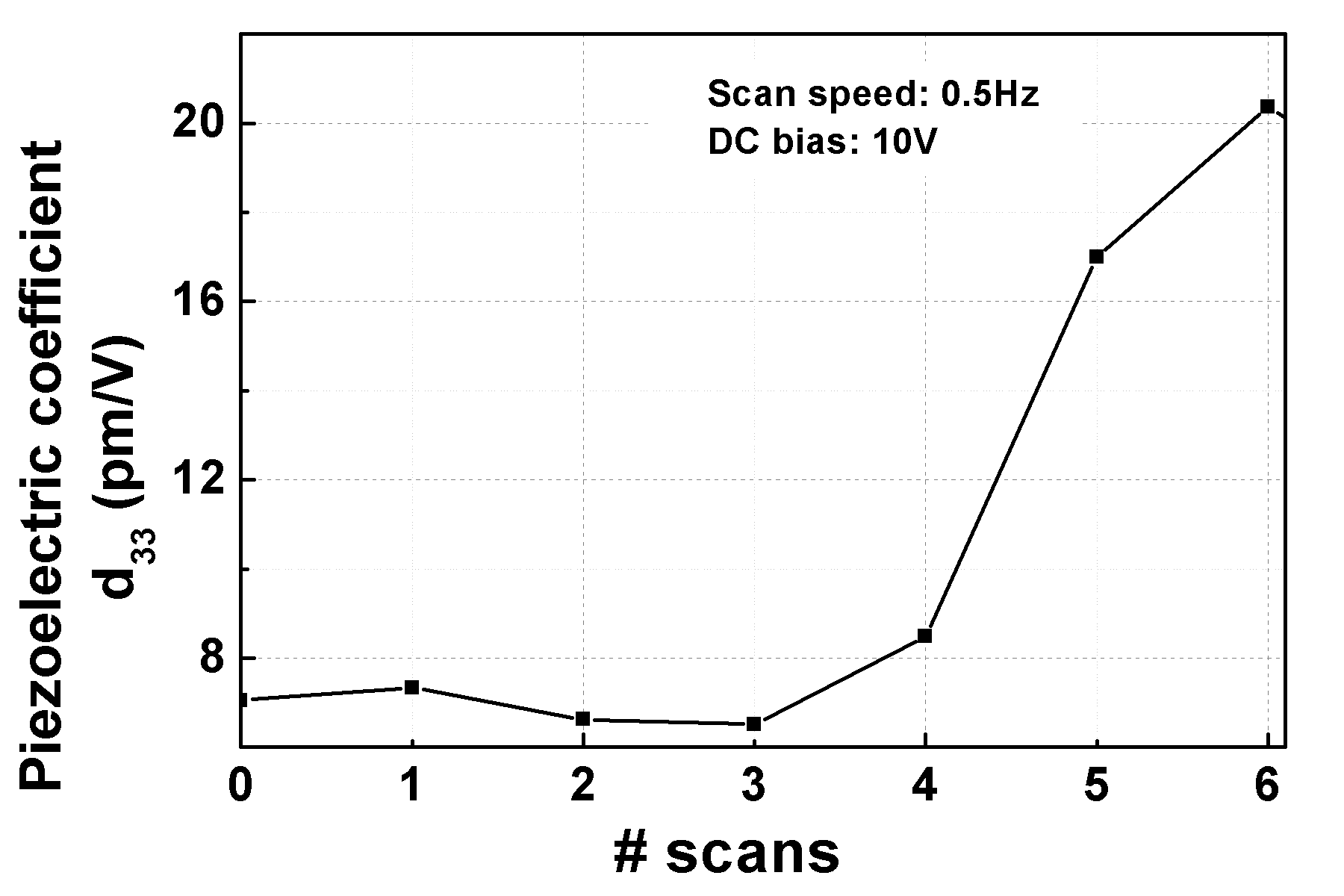}
   	\caption{The effective $d_{33}$ measured at the same location on the wafer increases with repeated number of scans in the c-AFM, indicating signs of polarization.}
   	\label{inconclusive}
   \end{figure}

\begin{figure}[htbp!]
      	\centering
      	\includegraphics[width= 9cm]{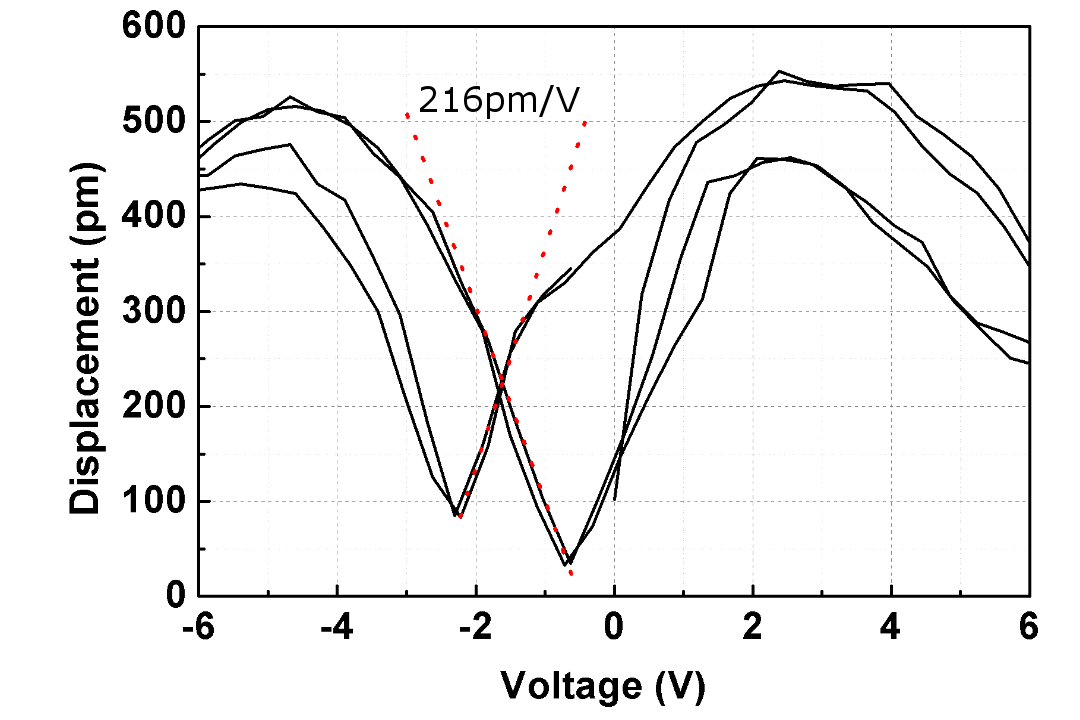}
      	\caption{Largest effective $d_{33}$ value recorded on the composite measured using PFM was $216pm/V$. The effective $d_{33}$ is obtained as slope of the linear region in the PFM response.}
      	\label{high_d33}
      \end{figure}

The sample prepared for determining the poling voltage using c-AFM consists of $1.7\mu m$ thick SU8/PMNPT composite film deposited on $20nm/200nm$ Cr/Au metal that is deposited on a silicon wafer,  without an intermediate oxide layer. Thus, the AFM tip serves as top electrode while scanning the sample, and the substrate contacted through the chuck forms the bottom electrode. Domain engineering is attempted by biasing the AFM tip and raster scanning across a $20\mu m$ x $20\mu m$ area on the film, following the process outlined in \cite{ryan}.
It should be noted that the piezoelectric coefficient varies greatly across the wafer and is measurable only directly atop the areas on the wafers that have nanoparticles, and thereby we are only able to report local measurements of $d_{33}$ on the wafer in regions where PMNPT nanoparticles are present. Fabricating a resonator that spans a large film area makes it possible to incorporate several nanoparticles in the transducer, and thus an effective boost in the transduction efficiency as compared to that offered by a single nanoparticle.
As seen in Figure \ref{inconclusive}, we observe that the effective $d_{33}$ measured locally at any point increases with increasing number of scans, thus indicating that the PMNPT particles can be polarized to improve the transduction efficiency. However we are unable to heat the sample beyond the Curie temperature of PMNPT in the AFM setup, and thus the entire film cannot be poled. The largest measured effective $d_{33}$ value obtained locally on our sample was $216pm/V$, as shown in Figure \ref{high_d33}.

  \subsection{Resonator characterization}  

  \begin{figure}[htbp!]
      	\centering
      	\includegraphics[width= 9cm]{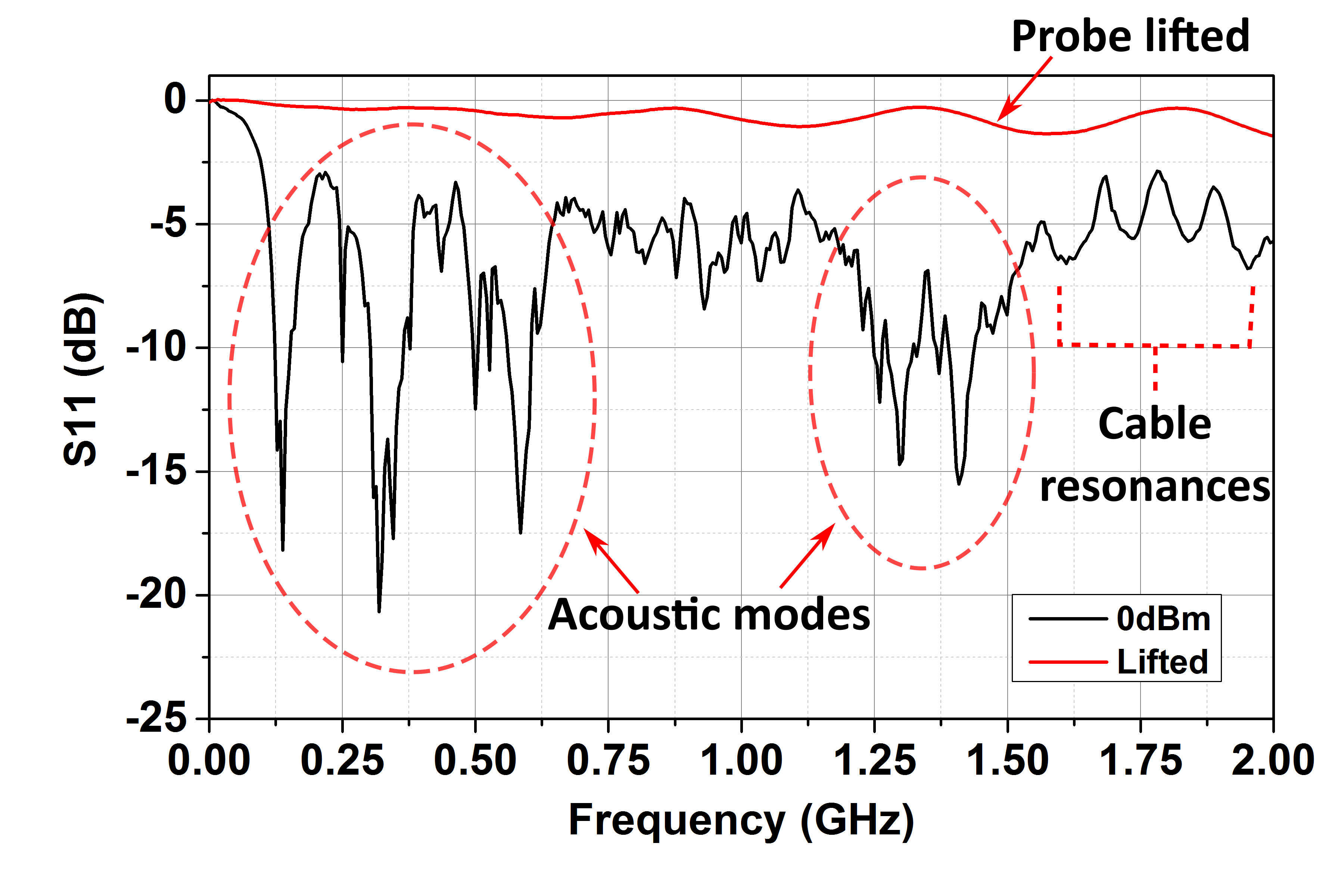}
      	\caption{Wide span measurement with network analyzer showing multiple acoustic resonances observed when sample is poled at $40V$ for 30 minutes. When the probe is lifted, the resonances are not observed, confirming their acoustic origin.
}
      	\label{S11}
      \end{figure}

      
We obtain the RF electromechanical spectrum of the HBAR devices using an Agilent E5071C Vector Network Analyzer (VNA) and GS probe tips. Due to unavailability of calibration substrates, we performed partial calibration of the cables using an SMA calibration kit. The DUT (device under test) is a circular shaped device with a radius of $250\mu m$ that was poled at $40V$ for $30$ minutes at ambient temperature using a Keithley 2604B DC source meter. After poling, the probes are connected to the VNA to record the 1-port reflection measurement $(S_{11})$ with applied RF power of $-10dBm$. Figure \ref{S11} shows multiple resonances observed in the device. We also measured the $S_{11}$ after lifting the RF probe, shown by the grey colored traces in Figure \ref{S11}, which shows cable resonances due to lack of full-calibration, but no device resonances. The center frequency of the HBAR resonances is estimated to be $c_{film}/2t_{film}=\frac{2490m/s}{2*1.7\mu m} \approx 730MHz$ where $c_{film}$ denotes longitudinal speed of sound in SU8 \cite{SU8_speed} and $t_{film}$ is thickness of the composite film. The frequency spacing is estimated to be $c_{sub}/2t_{sub}=\frac{8430m/s}{2*275\mu m} \approx 15.3MHz$, where $c_{sub}$ denotes longitudinal speed of sound in silicon and $t_{sub}$ is thickness of the $<$100$>$ silicon wafer. We observe multiple resonances in frequency range from $650MHz$ to $820MHz$, spaced uniformly by $\approx 15MHz$, however there are more efficiently transduced resonances at frequencies below $600MHz$ and between $1.1GHz$ to $1.5GHz$. We are further investigating the other resonances observed in the device and approaches to improve the transduction efficiency of the HBAR resonances.
      
\section{Conclusion and future work}
\begin{figure}[htbp!]
      	\centering
      	\includegraphics[width= 9cm]{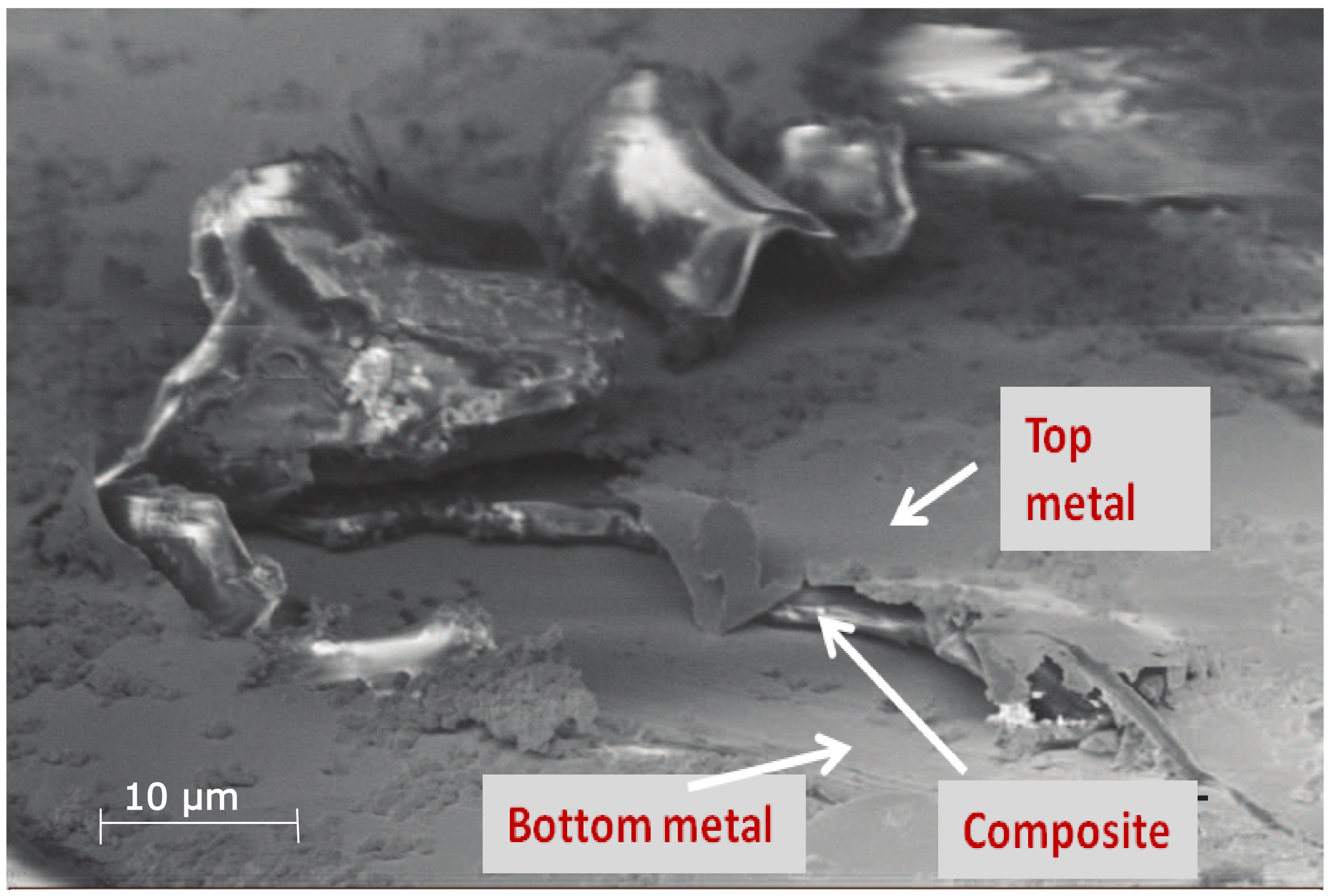}
      	\caption{SEM image showing cracks in the metal pad causing electrical short. The delamination of the piezoelectric film from the metals is also identified here as a reliability issue.}
      	\label{crack}
      \end{figure}

We present a novel piezoelectric-polymer composite based acoustic resonator that is processed entirely at room temperature. We provide a process flow to embed the piezoelectric material (PMNPT) into the polymer (SU8), and the resultant film is deposited via spin coating. We also successfully fabricated thickness mode resonators using the composite film on a low acoustic loss silicon wafer substrate, and experimentally confirmed the presence of acoustic resonances in the device. An unanticipated challenge encountered while performing the RF measurement was the cracking of the polymer film due to mechanical pressure applied upon landing the probe tips, thus causing electrical shorts between top and bottom electrodes. The adhesion of polymer matrix on metal surface is also identified as major issue from a reliability perspective. Upon inspection in SEM, we noticed the cracking of the polymer film and delamination of the metals, as shown in Figure \ref{crack}. The shorting can be addressed by incorporating a trench around the signal pad in future designs. In our current experiments, we take utmost care to land the electrodes gently, to not short out the DUT. The delamination warrants further investigation, that will be taken up in future work.
 
\section*{Acknowledgment}

The authors acknowledge the support of the Ministry of Electronics and Information Technology (MeitY) of the Government of India through the Centre of Excellence in Nanoelectronics (CEN) at IIT Bombay. We also thank Dr. Tanay Gosavi at Intel for engaging discussions on piezoelectric material characterization and HBAR fabrication; Ms. Shivangi Chugh and Dr. Siva Rama Krishna at IIT Bombay for help with RF testing; and Prof. Shalabh Gupta for access to RF measurement facilities. We also thank Mr. Ambika Shanker Shukla and IIT Bombay Nanofabrication Facility (IITBNF) staff for help with device fabrication.

\end{document}